\documentclass[aps,floatfix,nofootinbib,amsmath,amssymb]{revtex4}
\usepackage[dvips]{graphicx}

\begin{document}

\newcommand{\Frac}[2]{\frac
{\begin{array}{@{}c@{}}\strut#1\strut\end{array}}
{\begin{array}{@{}c@{}}\strut#2\strut\end{array}}}

%%%%%%%%%%%%%%%%%%%%%%%%%%

\title{Cosmology and the Unexpected\footnote{Two lectures presented at the
International School of Subnuclear Physics, \textit{Searching for the
`totally unexpected' in the LHC era,} Erice, Italy 2007.}}

\author{Edward W.\ Kolb}
\affiliation{Department of Astronomy and Astrophysics, Enrico Fermi Institute, 
and  Kavli Institute for Cosmological Physics, University of
Chicago, Chicago, IL \ \ 60637-1433 }

\begin{abstract}

In these two lectures I will discuss some outstanding problems in the
standard model of cosmology, concentrating on the physics that might
be related to the title of this school, ``Searching for the totally
unexpected in the LHC era.''  In particular, I will concentrate on
dark energy, dark matter, and inflation.

\end{abstract}

\maketitle

%%%%%%%%%%%%%%%%%%%%%%%%%%
%%%%%%%%%%%%%%%%%%%%%%%%%%
\section{Introduction}
%%%%%%%%%%%%%%%%%%%%%%%%%%
%%%%%%%%%%%%%%%%%%%%%%%%%%

In the last decade or two we have made remarkable progress in
measuring cosmological parameters to unprecedented
accuracy. Parameters such as the Hubble constant, $H_0=72\pm 8 \textrm{ km
s}^{-1}\textrm{ Mpc}^{-1}$
\cite{Hubblekey}, the temperature of the cosmic background radiation
(CBR), $T_0=2.728\pm 0.004\ \textrm{K}$ \cite{COBET}, and many other
parameters are now known to impressive precision.  Of course, we don't
invest so much time and effort to determine cosmological
parameters because of an interest in numerology, but rather, we do it 
because the parameters are necessary input for the
task of constructing a standard cosmological model.

\begin{figure}
\begin{center}
\includegraphics[width=11cm]{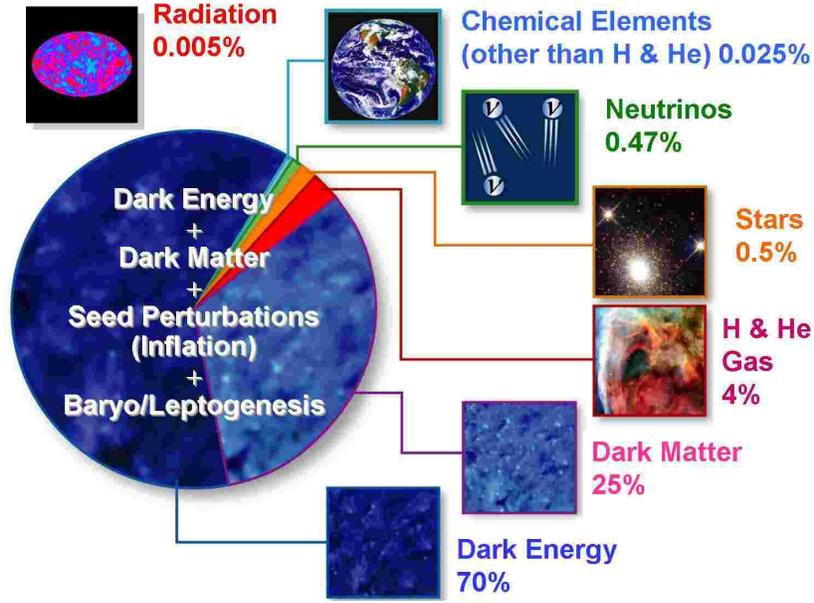}
\caption{Elements of the standard $\Lambda$CDM cosmological model.
Illustrated by the pie is the fraction of the mass-energy of the
Universe in various components.  Also noted are some other necessary
parts of the standard model, namely seed perturbations and
baryo/leptogenesis.}
\label{pie}
\end{center}
\end{figure}

The precision cosmological measurements have lead to the latest cosmological
model, usually called the \textit{standard cosmological model,} or
$\Lambda$CDM, where $\Lambda$ indicates Einstein's cosmological constant (or
more generally, dark energy), and CDM stands for cold dark matter.  Aspects of
this cosmological standard model relevant for the purposes of these lectures
are indicated in Figure \ref{pie}.  The most remarkable feature of the standard
cosmological model is that it seems capable of accounting for all cosmological
observations; i.e., it seems to work!

There are several interesting features of the $\Lambda$CDM model
illustrated in Fig.\ \ref{pie}. First, while the early universe was
radiation dominated, at present the radiation energy density is only
$0.005\%$ of the total.  The figure also illustrates why chemistry is
not very important: the chemical elements (by which I mean elements
other than the simplest elements of hydrogen and helium) are only about
$0.025\%$ of the total.  Neutrinos contribute a much larger fraction
of the total mass-energy density, about $0.5\%$.  The neutrino
contribution is about as large as the contribution from stars.  Most
of the baryons in the Universe are not found in stars, but rather they are
in the form of a hot intracluster gas of hydrogen and helium.  This
completes what we ``see'' (although there is only indirect evidence
for the cosmic neutrino background).

The most striking feature of the present composition of the universe
in the $\Lambda$CDM model is that $95\%$ of the present universe is
dark!  Of the dark components, roughly $25\%$ of the total mass-energy
density is dark matter, associated with galaxies, clusters of
galaxies, and other bound structures.  The bulk of the mass-energy of
the universe, about $70\%$ in the standard $\Lambda$CDM model, is dark
energy, which drives an accelerated expansion of the Universe.

In the first lecture I will discuss issues associated with dark energy.  In the
second lecture, I will discuss possibilities for dark matter.  Also in the
second lecture I will briefly discuss some aspects of cosmic inflation, in
particular what we might learn about inflation from present observations.

For the purposes of this school, let me begin by recalling Einstein's
comment to Arnold Sommerfeld on December 9, 1915, ``How helpful to us
is astronomy's pedantic accuracy, which I used to secretly ridicule.''
While Einstein was referring to precision measurements of the advance
of the perihelion of Mercury, now we find precision measurements
helpful to us because dark energy, dark matter, inflation, and
baryo/leptogenesis seem to point to physics beyond the standard model
of particle physics.

%%%%%%%%%%%%%%%%%%%%
%%%%%%%%%%%%%%%%%%%%%

%%%%%%%%%%%%%%%%%%%%%%%%%%
%%%%%%%%%%%%%%%%%%%%%%%%%%
\section{Dark Energy}
%%%%%%%%%%%%%%%%%%%%%%%%%%
%%%%%%%%%%%%%%%%%%%%%%%%%%

The most important point regarding dark energy, which I will stress
repeatedly, is that all evidence for dark energy is \textit{indirect.}
The only effect of dark energy is on the expansion history of the
Universe.

The expansion rate of the Universe is perhaps the most fundamental
quantity in cosmology.  In the standard cosmological model the
expansion rate is determined from the $00$-component of the Einstein
equations, $G_{00}=8\pi G T_{00}$.  In the
Friedmann-Robertson-Walker-Lema\^{i}tre (FLRW) model, $G_{00}=3H^2 +
k/a^2$, where $H=\dot{a}/a$ is the expansion rate (here, $a$ is the
Robertson-Walker scale factor and $k=\pm1$ or $0$ depending on the
geometry).  With the assumption of a perfect-fluid stress tensor,
$T_{00}=\rho$, where $\rho$ is the mass density.

The stress tensor is assumed to consist of several components with
different equations of state.  For any component $i$, the equation of
state is defined as $w_i=p_i/\rho_i$.  The energy density in component
$i$ evolves with redshift $z\equiv a_0/a-1$  as $\rho_i(z) =
\rho_i(0)\left(1+z\right)^{3(1+w_i)}$, where $a_0$ is the
present value of the scale factor.  In the event that $w$ is a
function of $a$, then $\left(1+z\right)^{3(1+w_i)}$ should be replaced
by $\exp\left\{3\int_a^{a_0} da\, a^{-1}\left[1+w(a)\right]\right\}$.  
Then the expansion rate as a function of redshift can be expressed as
\begin{equation}
H^2(z)=H_0^2\left[\left(1-\Omega_\textrm{TOTAL}\right)(1+z)^2 +
\Omega_M(1+z)^3 + \Omega_R(1+z)^4  +\Omega_w(1+z)^{3(1+w)}\right] .
\label{hsquared}
\end{equation}
Here, $\Omega_i$ is the ratio of the present energy density in
component $i$ compared to the critical density $\rho_C=3H_0^2/8\pi G$,
and $\Omega_\textrm{TOTAL}=\sum_i\Omega_i$.  The subscript ``$M$''
indicates a matter component ($w=0$) while the subscript ``$R$''
indicates a radiation component ($w=1/3$).

The value of $\Omega_\textrm{TOTAL}$ is well determined by WMAP to be
$1.026^{+0.015}_{-0.016}$ at the $68\%$ confidence level \cite{wmap3}.
The value of $\Omega_R$ is also well determined by CBR measurements:
It is of order $5\times 10^{-5}$.  Finally, $\Omega_w$ is determined
by measuring the expansion history of the Universe, and determining the
values of $\Omega_w$ and $w$ necessary to fit the data.

Many cosmological observables depend on the expansion history of the
Universe.  They often depend on the expansion history through the
coordinate distance $r$ of a source of redshift $z$.  Recall that the
Robertson-Walker metric can be written in the form
\begin{equation}
ds^2 = dt^2 - a^2(t)\left[\frac{dr^2}{1-kr^2}+r^2d\Omega^2\right] ,
\label{rw}
\end{equation}
where here $d\Omega$ is the angular differential and $r$ is the
comoving ``radial'' coordinate.  The radial coordinate of a source at
redshift $z$ is determined by integrating the null geodesic
($ds^2=0$) to obtain
\begin{equation}
r(z) = \left. \begin{array}{l} \sin \\ 1 \\ \sinh \end{array} \right\}
\left[ \int_0^z \frac{dz'}{H(z')} \right] ,
\label{rofz}
\end{equation}   
where $\sin$, $1$, $\sinh$ obtains for $k=+1,\ 0,  \-1$, respectively.

Now let us turn to the observables defined in Table \ref{obs}.  One
can see that they all depend on the time evolution of the expansion
rate, hence to the properties of the stress tensor, hence to a new
fluid characterized by $w$.

%%%%%%%%%%%%%%%%%%%%%%%%%%
\begin{table}
\caption{\label{obs} The relationship of observables to the time evolution of
the expansion rate of the Universe. The coordinate distance $r(z)$ depends on 
the time evolution of $H(z)$ through Eq.\ (\ref{rofz}).}
\begin{ruledtabular}
\begin{tabular}{lccr}
Observable  & Notation    &  Definition & Value   \\  \hline
luminosity distance & $d_L(z)$ & 
$(\textrm{Luminosity}/4\pi\textrm{Flux})^{1/2}$ & 
$d_L(z)\propto r(z)(1+z)$    \\
angular-diameter distance & $d_A(z)$ & 
$\textrm{Physical size}/\textrm{Angular size}$ &
$d_A(z)\propto r(z)/(1+z)$ \\ 
volume element & $dV(z)$ & 
$\sqrt{h}\, dr d\Omega$ &
$dV(z)=\Frac{r^2(z)}{\sqrt{1-kr^2(z)}}dr d\Omega $ \\
age of the Universe & $t(z)$ & 
time from $z=\infty$ to $z$ &
$t(z)=\int_z^\infty{\Frac{dz'}{(1+z')H(z')}}$  
\end{tabular}
\end{ruledtabular}
\end{table}
%%%%%%%%%%%%%%%%%%%%%%%%%%

\begin{figure}
\begin{center}
\includegraphics[width=15cm]{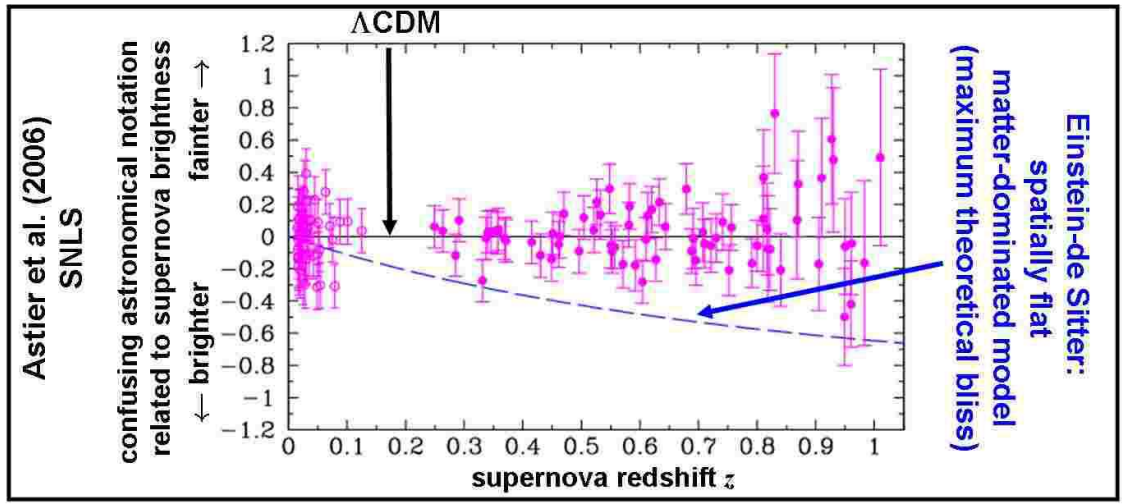}
\caption{The results of the Supernova Legacy Survey \cite{SNLS}. Shown
is the apparent magnitude of supernovae normalized to that expected in
the standard $\Lambda$CDM model.}
\label{sn1}
\end{center}
\end{figure}

The best evidence for dark energy comes from  luminosity-distance--redshift
determinations using Type 1a supernovae as standard candles of known (or at
least calibrated) luminosity \cite{riessperlmutter}.  A recent large
survey is the Supernova Legacy Survey (SNLS) \cite{SNLS}.  The results for the
observed magnitude of supernovae as a function of redshift is shown in Fig.\
\ref{sn1}. Without performing a statistical analysis, it is apparent that the
data fit the $\Lambda$CDM model and the Einstein--de Sitter (EdS) model is
``observationally challenged.''

%%%%%%%%%%%%%%%%%%%%
%%%%%%%%%%%%%%%%%%%%%

\begin{figure}
\begin{center}
\includegraphics[width=8cm]{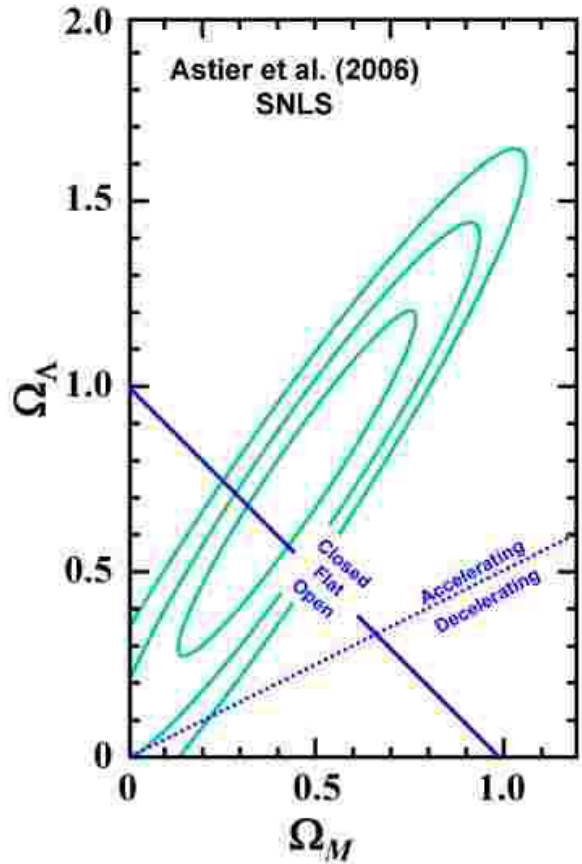}
\caption{Values of $\Omega_\Lambda$ and $\Omega_M$ allowed at
confidence levels of $68\%$, $90\%$, and $95\%$ assuming a
cosmological model with $w=-1$.  The data used to place the limits are
show in Fig.\ \ref{sn1}.}
\label{sn2}
\end{center}
\end{figure}

It is important to realize that dark energy experiments do not
actually measure dark energy directly.  They determine the expansion
history of the Universe by measuring cosmological observables
sensitive to the expansion history.  What is determined is that the
EdS model does not fit the data.  Any information about the new
component to the stress tensor depends on the assumptions
you put in.  You have to input a cosmological model and compare it to
the data.  An example of such an analysis is shown in Fig.\ \ref{sn2}.
To construct this figure, it was assumed that the expansion history of
the Universe is described by a FLRW model with dark energy described
by a fluid with $w=-1$, i.e., Einstein's cosmological constant.  Also
employed are priors on cosmological parameters such as $H_0$.

%%%%%%%%%%%%%%%%%%%%
%%%%%%%%%%%%%%%%%%%%%

If one adopts the model used in Fig.\ \ref{sn2}, then there is a
cosmological constant with an equivalent energy density of $\rho_V\sim
10^{-30} \textrm{g cm}^{-3}$.  Henceforth I will refer to this as the
\textit{cosmoillogical} constant.  There are two reasons for referring
to this cosmo constant as illogical, rather than logical.

First, the magnitude of the cosmoillogical constant is, well, illogical.
Length scales and mass scales of the cosmoillogical constant are given in
Table \ref{illogical}.  Fundamental length scales and mass scales
shown in Table \ref{illogical} are, as I said, illogical.

%%%%%%%%%%%%%%%%%%%%%%%%%%
\begin{table}
\caption{\label{illogical} The length scales and mass scales implied
by a cosmoillogical constant with an energy density of
$10^{-30}\textrm{g cm}^{-3}$.}
\begin{ruledtabular}
\begin{tabular}{lcr}
             & $\rho_V$    &  $\Lambda=8\pi G$    \\  \hline
length scale $(\rho_V)^{-1/4}$ & $10^{-3}\textrm{cm}$ & 
$10^{29}\textrm{cm}$  \\
mass scale   $(\rho_V)^{1/4}$  & $10^{-4}\textrm{eV}$ & 
$10^{-33}\textrm{eV}$  
\end{tabular}
\end{ruledtabular}
\end{table}
%%%%%%%%%%%%%%%%%%%%%%%%%%

A cosmoillogical constant is equivalent to, and indistinguishable from,
contributions to the vacuum energy.  It is only meaningful to consider
them together as a vacuum energy.  Among the contributions to vacuum
energy is the fact that all fields are harmonic oscillators with a
zero point energy.  This should contribute a vacuum energy of
$\rho_V=\sum_\textrm{all particles} \pm \int d^3k\ \sqrt{k^2+m^2}$,
where the sign of the contribution depends on the spin-statistics of
the particle.  The individual integrals have quartic divergences, so
formally the answer is infinity. (Unless for some reason--such as
supersymmetry (SUSY)--there are exact cancellations between bosons and
fermions.) This need not cause alarm because, after all, this is field
theory.  One might imagine introducing a cutoff $\Lambda_C$ to the
integral so that the integral is finite, and the individual contributions
are $\pm \Lambda_C^4$.  If you assume $\Lambda_C$ is related to
gravity, then this contribution to $\rho_V$ is
$M_{Pl}^4=(10^{28}\textrm{ eV})^4=10^{112}\textrm{ eV}^4$.  If you
conjecture that $\Lambda_C$ is related to the SUSY breaking scale
(assumed here to be $1$ TeV), then $\rho_V = M_\textrm{SUSY}^4 =
(10^{12}\textrm{ eV})^4=10^{48}\textrm{ eV}^4$.  These scales are far
away from the observed value of $\rho_V=(10^{-4}\textrm{
eV})^4=10^{-16}\textrm{ eV}^4$.

Another contribution to vacuum energy comes from the vacuum energy
associated with spontaneous symmetry breaking.  Associated with
spontaneous symmetry breaking of GUTs, SUSY, electroweak, and chiral
symmetries are vacuum energies of $10^{100}\textrm{eV}^4$,
$10^{48}\textrm{eV}^4$, $10^{45}\textrm{eV}^4$, and
$10^{32}\textrm{eV}^4$, respectively.  Once again, these scales are
far away from the observed value of $\rho_V$.

\begin{figure}
\begin{center}
\includegraphics[width=11cm]{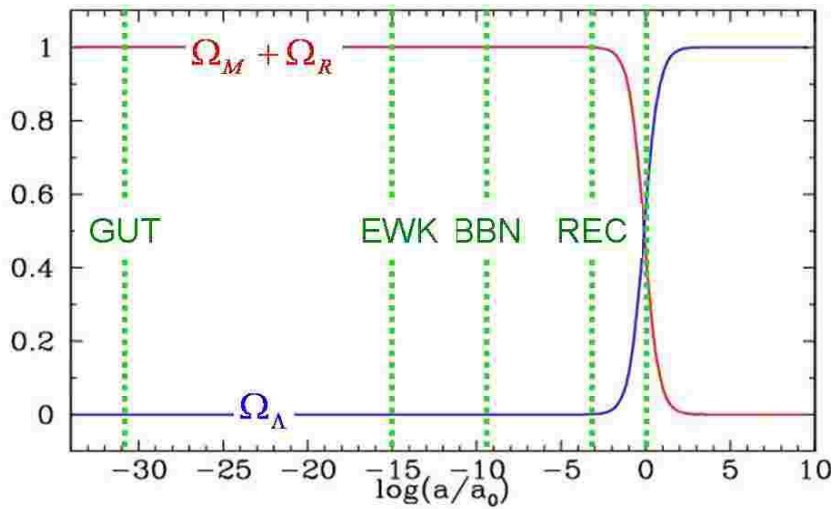}
\caption{The relative contributions of $\Omega_M+\Omega_R$ and
$\Omega_\Lambda$ as a function of the scale factor $a$.}
\label{timing}
\end{center}
\end{figure}

A not unrelated issue is the timing of the epoch in the history of the
Universe when the dark energy density is comparable to the density of
matter and radiation.  A graph of the relative contributions of
$\Omega_M+\Omega_R$ and $\Omega_\Lambda$ is shown in Fig.\
\ref{timing}.  It turns out to be very curious that while in the
distant past vacuum energy was ``in the noise,'' and in the
distant future matter and radiation densities will be ``in the
noise,'' while today we live at the very convenient epoch when they are just
about the same value.  More than curious, this fact is illogical.

%%%%%%%%%%%%%%%%%%%%
%%%%%%%%%%%%%%%%%%%%%

Astronomical evidence for a vacuum energy today would be
remarkable.  But as the saying goes, remarkable results require
remarkable evidence.  The evidence from supernova observations has
been tested and probed for systematic uncertainties.  While there is
no indication of anything fishy, it is crucial to find
corroborating evidence from other techniques with different
astronomical systematic uncertainties.  Fortunately, there is other
evidence that there is dark energy.  (As discussed below, a more exact
statement is that there is other evidence that the time evolution of
the expansion rate is not described by the EdS model, and well fit by
$\Lambda$CDM.)

Other evidence comes from 1) cosmic subtraction, 2) baryon acoustic
oscillations, 4) weak lensing, 4) galaxy clusters, 5) the age of the
Universe, and 6) structure formation.

\begin{figure}
\begin{center}
\includegraphics[width=11cm]{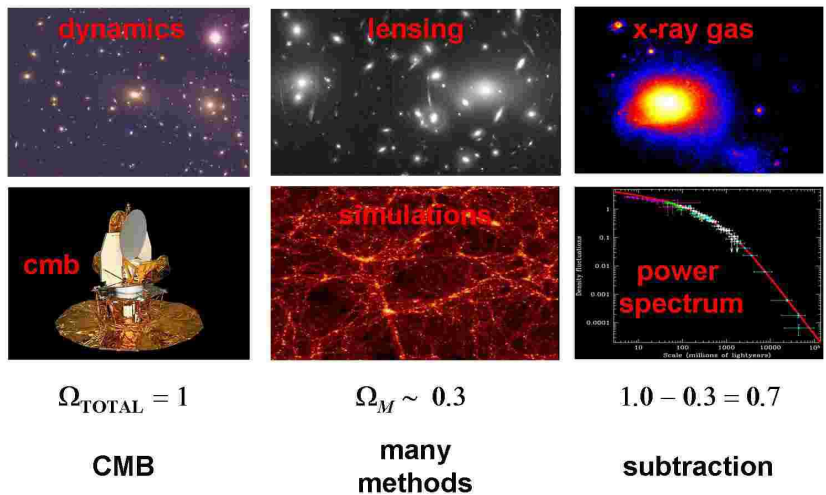}
\caption{Cosmic subtraction.  CBR observations imply
$\Omega_\textrm{TOTAL}\simeq 1$, while a variety of techniques imply
$\Omega_B\simeq 0.3$. The remaining $0.7$ is readily attributed to
dark energy.}
\label{subtraction}
\end{center}
\end{figure}

First let me mention the sophisticated mathematical operation of
cosmic subtraction.  The idea is illustrated in Fig.\
\ref{subtraction}. The idea is simple.  CBR observations imply
$\Omega_\textrm{TOTAL}$ is very near unity.  A variety of astronomical
observations illustrated in Fig.\ \ref{subtraction} result in
$\Omega_M\simeq0.3$.  (The observations resulting in this value of
$\Omega_M$ include dynamics of galaxies and galaxy clusters,
gravitational lensing, X-ray observations of galaxy clusters that
measure the depth of the potential well by determining the gas
temperature, CBR observations, numerical simulations of the
large-scale structure of the Universe, and determination of the
``shape factor'' of the power spectrum of density fluctuations.)

Now for the promised sophisticated operation of subtraction:
\begin{equation} 1-0.3 = 0.7. \end{equation} Let me expand on this for
the benefit of any string theorists who might be in the audience.
$1-0.3$ is not equal zero, but in fact, it is $0.7$.  This profound
result holds in any spacetime dimension and is perfectly consistent
with the AdS/CFT correspondence.  This ``missing'' component to the
total stress energy of the Universe is conveniently attributed to dark
energy.

%%%%%%%%%%%%%%%%%%%%
%%%%%%%%%%%%%%%%%%%%%

Before turning to other techniques, it is a good time to re-emphasize the
important point made previously.  The measurements \textit{do not} tell us that
there is dark energy.  Rather, the observations only provide indirect evidence
for dark energy.  To elaborate on this, let me review the procedure.
\begin{enumerate}
\item Assume a model cosmology.  In this case, the assumptions involve:
	\begin{enumerate}
	\item The FLRW model, which results in the Friedmann equation for 
	$H(z)$.
	\item The energy and pressure content of the Universe and how
	they scale with redshift.
	\item Input or integrate over cosmological parameters such as $H_0$,
	$\Omega_B$, etc.
	\end{enumerate}
\item Within the cosmological model, calculate observables such as $d_L(z)$,
$d_A(z)$, $H(z)$, etc.
\item Compare to observations.
\item Discover that the model cosmology fits with dark energy, but not
without dark energy.
\end{enumerate}

So what the observations seem to tell us is that the observed $H(z)$
is not described by $H(z)$ calculated from the EdS model, but is well
described by $H(z)$ described by the $\Lambda$CDM model.

Again, all evidence for dark energy is indirect!

So, what should we do?  We can no longer just hide from the data.  The
evidence is overwhelming that $H(z)$ is not described the by the EdS
model.  Again, this is not just from the supernova projects, but from
a variety of techniques.

So we don't know what the answer is, but we do seem to know something
very valuable: We know an equation that doesn't work; namely,
\begin{equation}
G_{00}(\textrm{FLRW}) \neq 8\pi G T_{00}(w=0\textrm{ perfect fluid matter}).
\label{nonequation}
\end{equation}
Let's exploit this non-equation and imagine how we can fix it.  

First, we might try to add something to the right-hand side of the
equation; a $\Delta T_{00}$ if you will.  This $\Delta T_{00}$ is
usually what we mean by dark energy.

Theorists employ two tools in constructing a $\Delta T_{00}$.  Now
experimentalists believe that theorists don't know anything about
tools, but that is not true.  In my own case, I owned a British sports
car for over 20 years (a 1965 Austin-Healey Sprite) and an Italian
economy car (a 1974 Fiat 128) for more than 5 years.  Anyone who has
owned either of those marvels of engineering has had to use tools very
often.  In my experience, you only need two tools to fix anything:
duct tape and WD-40.  There are only two rules:
\begin{enumerate}
\item If something moves and it shouldn't, use duct tape.
\item If something doesn't move and it should, squirt it with WD-40.
\end{enumerate}

The equivalent theoretical tools used for dark energy are the anthropic (or
Landscape is you speak string) explanation,\footnote{The anthropic
explanation is sometimes referred to as the anthropic
\textit{principle.}} and scalar fields (known as quintessence for this
purpose).

The anthropic/Landscape/duct-tape idea is that there are many sources
of vacuum energy.  String theory is conjectured to have many (perhaps
more than $10^{500}$!) vacua.  The different vacua have different
values of the total vacuum energy density.  Presumably most of them
have a vacuum energy much, much larger (say, by a factor of $10^{100}$)
than the observed value.  But don't worry, since although
exponentially uncommon, vacua with vacuum energy as small as observed
are hospitable for life while the more common values result in an
inhospitable universe.  The anthropic idea is that we should only 
consider cosmological models with values of the vacuum energy that are
hospitable for life. Therefore, there is an anthropic selection effect.

So you see that the anthropic explanation is just like duct tape.  There
were some loose ends after a calculation.  Like duct tape, the
anthropic explanation need not be elegant or permanent, but it ties
down the loose ends.

Now the quintessence idea borrows a concept from high-energy physics:
whenever there is dynamics that is poorly understood, invent a scalar
field.\footnote{One might think the fact that there is no known
fundamental scalar field might temper the enthusiasm for introducing
them.}  One way to visualize the quintessence field is that there are
many contributions to $\rho_V$.  Some unknown principle or dynamics
results in the minimum value $\rho_V=0$, but we are not there yet--we
are still evolving to the ground state.

Now let's turn to ideas for modifying the left-hand side of
non-equation (\ref{nonequation}), adding a $\Delta G_{00}$.  The first
class of ideas involves modifying gravity.  The first 
type  in this class involves extra dimensions.
\begin{enumerate}
\item The braneworld modifies the Friedmann equation \cite{BDL}.  In
some models with branes the Friedmann equation does \textit{not} arise
from the $00$-component of the Einstein equation, but rather from the
Israel discontinuity condition as one crosses the brane.
\item In some extra-dimension models the gravitational force law is
modified at \textit{large} distances \cite{DDG}.
\item In models with branes the wavefunction of gravitons can leak into the
bulk \cite{GRS,DGP}.  This is a sort of ``tired graviton'' explanation for 
dark energy.
\item There are extra-dimension theories where gravity becomes repulsive at
$R\sim\textrm{Gpc}$ \cite{CEHT}.
\item All models of extra dimensions have ``Kaluza--Klein'' (KK)
excitations.  It could be that the $n=1$ KK excitation of the graviton
is very light, say with a mass $m\sim (\textrm{Gpc})^{-1}$
\cite{KMPRS}.
\end{enumerate}

Then there are theories (not necessarily based on extra dimensions) of
the type that modify the low-energy gravitational action.  The idea is
that Einstein and Hilbert got it wrong in 1915, and rather than the
Einstein-Hilbert action for gravity, $S=(16\pi G)^{-1}\int
d^4x\sqrt{-g}R$, it is more general, say of the form $S=(16\pi
G)^{-1}\int d^4x\sqrt{-g}f(R)$, where $f(R)$ is some function of, for
instance, the Ricci scalar $R$ \cite{CDTT}.  While it is very
reasonable to expect terms like $R^2$ or $R^{\mu\nu}R_{\mu\nu}$, which
affect the theory in the ultraviolet, what seems to be required for
dark energy are terms that affect the theory in the infrared.  Terms
like $R^{-1}$ would work, but they lead to problems.

One of the things we have come to understand from all these approaches
is how difficult it is to modify gravity on large scales without
destroying agreement with observations on solar system scales, or
leading to non-linearities, ghosts, or other theoretical
unpleasantness.

Finally, there is the idea (very unpopular in most quarters), that the
Friedmann equation has large corrections due to inhomogeneities in the
Universe \cite{R,KMNR,N,KMR}.  This idea is still under development,
but it is my favorite idea (since I have worked on it a lot).

Perhaps theoretical physicists should turn to astronomers and use the
famous quote of Einstein, ``Nothing more can be done by the
theorists. In this matter it is only you, the astronomers, who can
perform a simply invaluable service to theoretical
physics.''\footnote{Einstein wrote this in August 1913 to Berlin
astronomer Erwin Freundlich, encouraging him to mount a solar-eclipse
expedition to measure the bending of starlight as it passed near the
sun.  The astronomer eagerly accepted the challenge from the
theoretical physicist.  Unfortunately for the astronomer, the eclipse
of 1914 was in the Crimea during the outbreak of the First World War.
In an extraordinary rendition, Freundlich was captured, his equipment
confiscated, and he was imprisoned as an enemy combatant.  Eventually
he was released, but of course he missed the eclipse. This is just as
well, because in 1914 Einstein's prediction for the deflection of
light by the sun on the basis of his (at the time) incomplete theory
of gravity was wrong}

So now let us turn to astronomy and take a look at strategy and
techniques for dark energy.  Recently in the U.S., the Dark Energy
Task Force (DETF) \cite{DETF} was commissioned by the Department of
Energy, NASA, and the NSF to recommend strategy and approaches to
study dark energy.

The strategy recommended by the DETF is
\begin{enumerate}
\item Determine as well as possible whether the expansion is
consistent with being due to a cosmoillogical constant ($w=-1$).
\item If the expansion is not due to a cosmoillogical constant, probe
as well as possible the underlying dynamics by measuring as well as
possible the time evolution of dark energy, i.e., determine $w(a)$.
\item Search for a possible failure of general relativity through comparison of
the effect of dark energy on cosmic expansion with the effect of dark energy on
the growth of cosmological structures like galaxies or galaxy clusters.  
\end{enumerate}

%%%%%%%%%%%%%%%%%%%%
\begin{figure}
\begin{center}
\includegraphics[width=11cm]{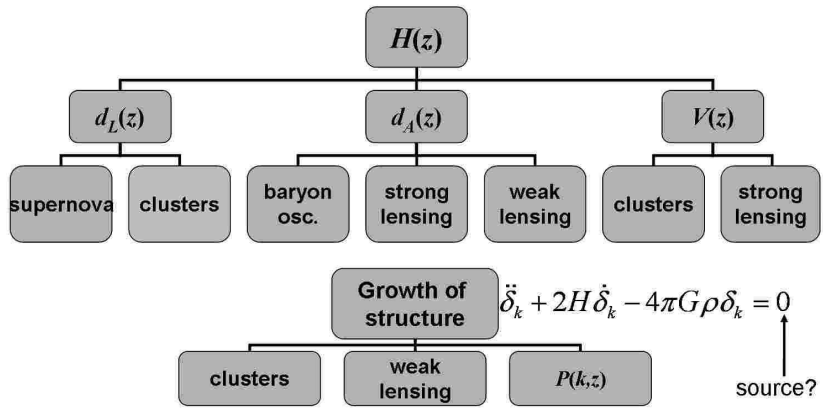}
\caption{Elements of a dark-energy program.}
\label{program}
\end{center}
\end{figure}
%%%%%%%%%%%%%%%%%%%%%

An observation program to accomplish this strategy is illustrated in
Fig.\ \ref{program}.  The effect of dark energy on $H(z)$ is probed by
observables like $d_L(z)$, $d_A(z)$, and $V(z)$.  Measurements of
these observables can be done through measurements of supernovae,
galaxy clusters, baryon acoustic oscillations, and lensing.
Observations of the growth of structure is interesting because the
expansion rate $H(z)$ enters the equation for the growth of
instabilities (in the linear regime the equation is given in the
figure).  If gravity is modified there is potentially a source term
that must be included.  

The DETF identified four main techniques for studying dark energy by
determining the time evolution of $H(z)$: 1) supernovae, 2) baryon
acoustic oscillations, 3) weak lensing, and 4) galaxy cluster surveys.
I will say a few words about each technique.

\subsection{Supernovae}

As you can see from Fig.\ \ref{program}, we use supernovae to study
dark energy by using them as standard candles and measuring the
luminosity-distance as a function of redshift.  This is a well
developed and well practiced technique, so I won't say very much about
it. Unlike the techniques discussed below, with supernovae there is a
lot of information per object.  Unclear at present is how good a
standard candle supernovae can be after calibration.  The fact that we
do not really \textit{know} how Type Ia supernovae actually explode
might intimidate the timid, but luckily astronomers are not timid.  It
is unclear how well the supernova technique will prove to be.  There
is a lot of discussion and debate about whether it is necessary to
make space observation, and whether photometric
redshifts\footnote{Redshifts traditionally have been taken by
spectroscopic techniques.  Photometric redshifts (photo-$z$'s) use
multicolor photometry as a crude spectrograph to determine the
redshift.} will be useful.

Since the supernova technique is so well known and familiar to most
people, I won't review it further, but use the time to discuss less
familiar techniques.

%%%%%%%%%%%%%%%%%%%%%%%%%%
%%%%%%%%%%%%%%%%%%%%%%%%%%
\subsection{Baryon Acoustic Oscillations}
%%%%%%%%%%%%%%%%%%%%%%%%%%
%%%%%%%%%%%%%%%%%%%%%%%%%%

\begin{figure}
\begin{center}
\includegraphics[width=11cm]{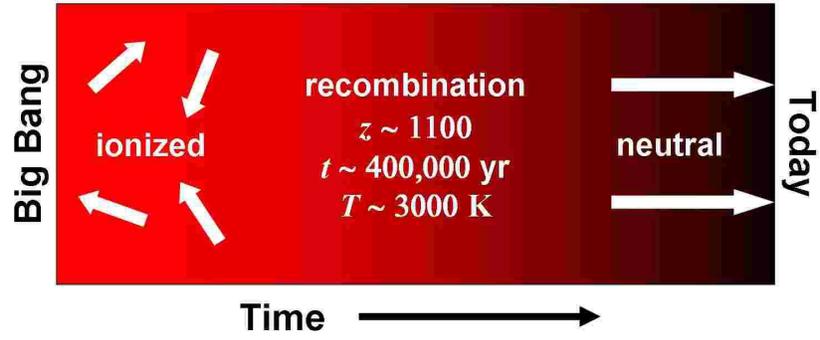}
\caption{A cartoon describing the transition between photons and
baryons tightly coupled before recombination in the early universe,
and photons free from baryons after recombination in the late
universe.}
\label{bao1}
\end{center}
\end{figure}

The baryon acoustic oscillations (BAO) technique is less familiar, so
I will discuss a little bit of the physics behind it.  It might be
useful to refer to Fig.\ \ref{bao1} during the discussion.

%%%%%%%%%%%%%%%%%%%%
%%%%%%%%%%%%%%%%%%%%%

Before recombination, the Universe was ionized.  The photons provided
an enormous pressure and restoring force preventing baryons from
moving.  Any perturbations in the baryons would not grow, but would
oscillate as sound waves.

After recombination, the universe is neutral and photons can travel
freely.  The baryon perturbations can grow or fall into dark matter
potential wells.

%%%%%%%%%%%%%%%%%%%%
\begin{figure}
\begin{center}
\includegraphics[width=11cm]{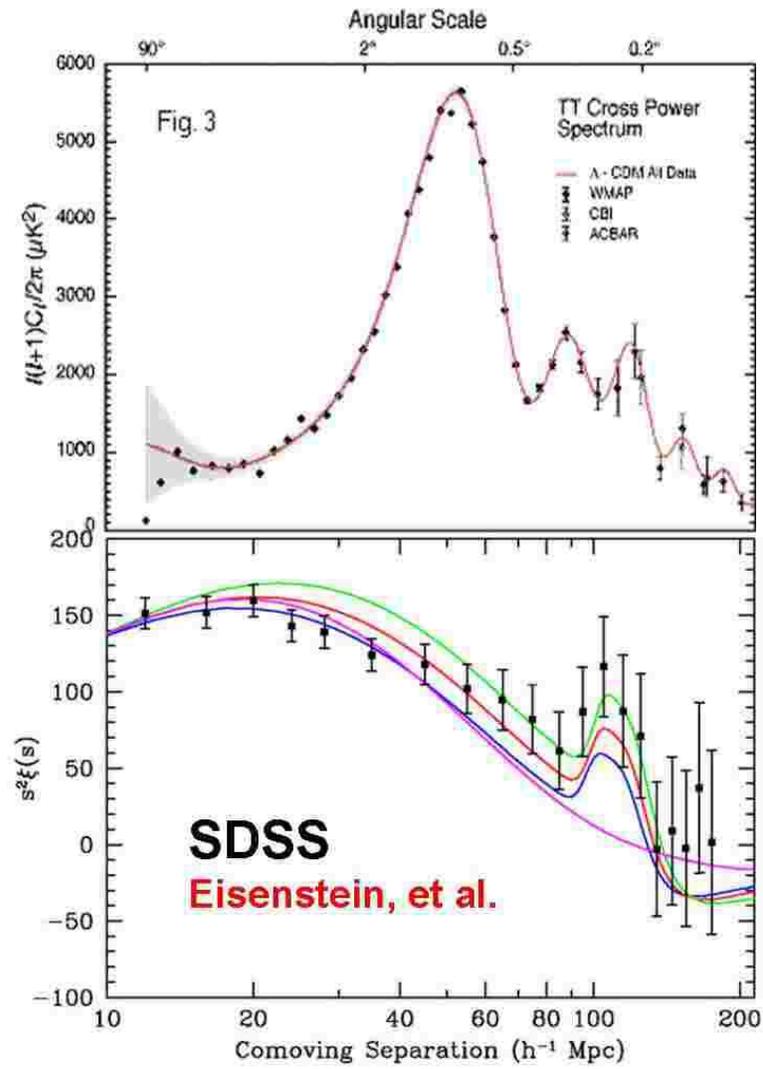}
\caption{The acoustic peaks in the angular power spectrum of CBR
fluctuations \cite{wmap3} and the BAO feature in the matter two-point
correlation function \cite{sdssbao}.}
\label{bao2}
\end{center}
\end{figure}
%%%%%%%%%%%%%%%%%%%%%

The phenomena of interest for BAO occurs in the transition between the
tightly coupled and decoupled regimes.  As recombination starts to
occur, the photon mean-free-path becomes long.  While the baryons are
still coupled to the photons, as the photons stream out of
overdensities, they drag the baryons along.  (This is known as Silk
damping.)  You can view this by considering an initial overdensity of
a mixture of dark matter, photons, neutrinos, and baryons.  When
recombination begins, the overdensity launches a spherical shock wave
in the photon-baryon fluid traveling outward at a velocity of
$c/\sqrt{3}$.  Eventually, the photons completely decouple from the
baryons, the baryons loose their pressure support, and the sound speed
of the baryon shock wave plummets.  Eventually, the shock stalls after
traveling a distance of about $150\textrm{ Mpc}$.  What is left behind
is an overdensity in the baryons on a distance scale of $150\textrm{
Mpc}$.  Dark matter falls into the baryon overdensities and the net
effect is a feature in the two-point correlation function of density
perturbations on a scale of $150\textrm{ Mpc}$.

This physics behind this feature in the matter distribution is exactly
the physics behind the acoustic peaks in the angular power spectrum of
CBR fluctuations.  This is illustrated in Fig.\ \ref{bao2}.

So nature provides us with a standard ruler of $150\textrm{ Mpc}$.
This can be used to probe dark energy through measuring its angular
size as a function of redshift, $d_A(z)$, and by measuring its radial
size, $H^{-1}\delta(z)$.

This technique has several advantages: The physical size of the
feature can be determined by the well measured CBR acoustic peaks.  It
is purely a geometric technique, largely free of astrophysical
systematic errors.

The technique has a couple of disadvantages: The feature in the power
spectrum is small (this is because $\Omega_B \ll \Omega_M$).  Since
the fundamental scale is $150\textrm{ Mpc}$, huge survey volumes are
required to include many, many $150\textrm{ Mpc}$ scales.  Finally,
nonlinear effects are important at smallish $z$, so the signal is
cleaner at $z\sim 2$ to $3$.  However, if dark energy behavior with $z$
is close to a constant, the sweet spot for dark energy is at lower
redshifts.

BAO is an emerging technique, largely free of systematic errors, with
promise for the future.

%%%%%%%%%%%%%%%%%%%%%%%%%%
%%%%%%%%%%%%%%%%%%%%%%%%%%
\subsection{Weak Lensing}
%%%%%%%%%%%%%%%%%%%%%%%%%%
%%%%%%%%%%%%%%%%%%%%%%%%%%

\begin{figure}
\begin{center}
\includegraphics[width=11cm]{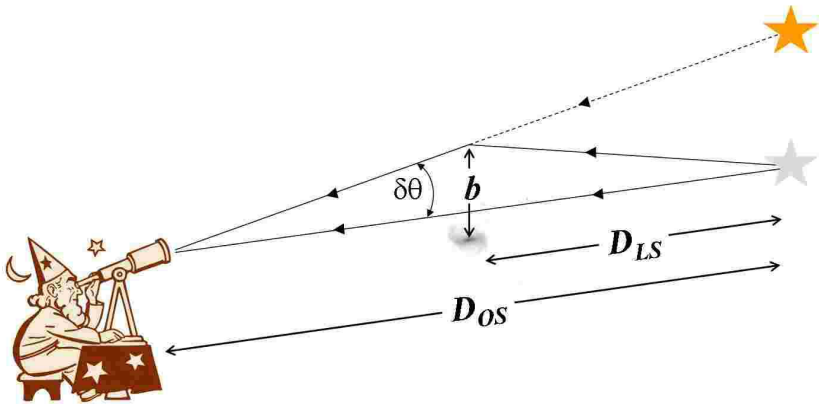}
\caption{A cartoon illustrating observation of lensing.}
\label{wl1}
\end{center}
\end{figure}

The weak lensing technique is illustrated in Fig. \ref{wl1}.  Suppose
there is a source at a distance of $D_{OS}$ and a gravitational lens
at a distance from the source of $D_{LS}$.  If the impact parameter is
$b$, the deflection angle is
\begin{equation}
\delta\theta=\frac{4GM}{b}\frac{D_{LS}}{D_{OS}} .
\label{deltatheta}
\end{equation}

Dark energy enters the game because geometric distances $D_{LS}$ and
$D_{OS}$ are affected by the time evolution of $H(z)$.  Dark energy
also affects the growth rate of the lensing mass $M$.

Generally, the deflection angles are so small that the signal for any
given galaxy is very small.  Weak lensing is based on the idea that
gravitational lensing leads to a distortion (or shearing) of the
source image.  Although the signal per galaxy is small, there are a
\textit{lot} of galaxies!  Since huge numbers of galaxies are
required, it is likely that photo-$z$'s will have to be used.

%%%%%%%%%%%%%%%%%%%%%%%%%%
%%%%%%%%%%%%%%%%%%%%
%%%%%%%%%%%%%%%%%%%%%
%%%%%%%%%%%%%%%%%%%%%%%%%%

There are two main sources of systematic errors involved in the weak
lensing technique.  The first is the uncertainty in photo-$z$'s.  The
second is related to the point spread function of the telescope.
Effects of the atmosphere and imperfections in the optics lead to a
distortion of images unrelated to lensing.  This leads to a debate
about whether weak lensing should be done from space or from ground.
Space observatories have the advantage of no atmosphere to distort the
image, and the ability to push into the near-infrared where errors in
photometric redshifts are expected to be less. Ground observatories
have the advantage of a larger aperture and less expensive telescopes.

The weak lensing landscape now has current projects imaging hundreds
of square degrees with deep multicolor data and thousands of square
degrees in shallow two-color data. In the near future, projects such
as the Dark Energy Survey will image thousands of square degrees with
deep multicolor data, and eventually LSST will image an entire
hemisphere very deep in six colors.

Weak lensing is also an emerging technique, with great promise.

%%%%%%%%%%%%%%%%%%%%%%%%%%
%%%%%%%%%%%%%%%%%%%%%%%%%%
\subsection{Galaxy Clusters}
%%%%%%%%%%%%%%%%%%%%%%%%%%
%%%%%%%%%%%%%%%%%%%%%%%%%%

Galaxy cluster surveys measure galaxy cluster masses, redshifts, and
spatial clustering.  They are sensitive to dark energy through the
volume--redshift relation, the angular-diameter--redshift relation,
the growth rate of structure, and the amplitude of clustering.

In my opinion, the advantage of the cluster technique is that it is
sensitive to dark energy in many ways.  There are problems associated
with the cluster technique.  The cluster selection must be well
understood, it is unclear what proxy should be used to determine the
cluster mass, and photo-$z$'s will likely have to be used.

Galaxy clusters are yet another emerging technique with promise.

%%%%%%%%%%%%%%%%%%%%%%%%%%
%%%%%%%%%%%%%%%%%%%%%%%%%%
\subsection{Combining Techniques}
%%%%%%%%%%%%%%%%%%%%%%%%%%
%%%%%%%%%%%%%%%%%%%%%%%%%%

In my opinion there is no single technique that should be pursued in
exclusion of the others.  There are three reasons for this opinion.
First, we will soon be at the point where systematic errors will
dominate the uncertainties. Different techniques have different systematic
uncertainties.  Second, we have no idea about the
nature of dark energy, and it is important to see its effects in
several ways.  For instance, if a new gravity theory is the answer, it
is possible that the effect on the growth of structures may be
different than the effect on luminosity distances.  Finally, different
techniques have different degeneracies with other cosmological
parameters, and great improvements in accuracy can be gained by
combining techniques \cite{DETF}.

%%%%%%%%%%%%%%%%%%%%%%%%%%
%%%%%%%%%%%%%%%%%%%%%%%%%%
\section{Dark Matter}
%%%%%%%%%%%%%%%%%%%%%%%%%%
%%%%%%%%%%%%%%%%%%%%%%%%%%

Now let me turn to the issue of dark matter. Refer back to Fig.\
\ref{subtraction}.  There are several methods to determine
$\Omega_M$. They all point to a value of $\Omega_M\sim 0.3$.  There
are also many reasons to believe that the baryon contribution
$\Omega_B$ is about $0.04$.  The two best pieces of evidence for
$\Omega_B$ are big-bang nucleosynthesis (BBN) and CBR
observations. Now that we have mastered subtraction, we can see that
the bulk of the matter density is non-baryonic. Here, I wish to
discuss possibilities for dark matter, so I won't spend a lot of time
motivating the fact that there must be dark matter.  I will also
assume that dark matter is non-baryonic.

The list of non-baryonic dark matter candidates is quite long.  An
entire lecture could be devoted just to listing them.  Rather than do
that, I will just concentrate on a few possibilities.

%%%%%%%%%%%%%%%%%%%%%%%%%%
%%%%%%%%%%%%%%%%%%%%%%%%%%
\subsection{Cold Thermal Relic}
%%%%%%%%%%%%%%%%%%%%%%%%%%
%%%%%%%%%%%%%%%%%%%%%%%%%%

\begin{figure}
\begin{center}
\includegraphics[width=11cm]{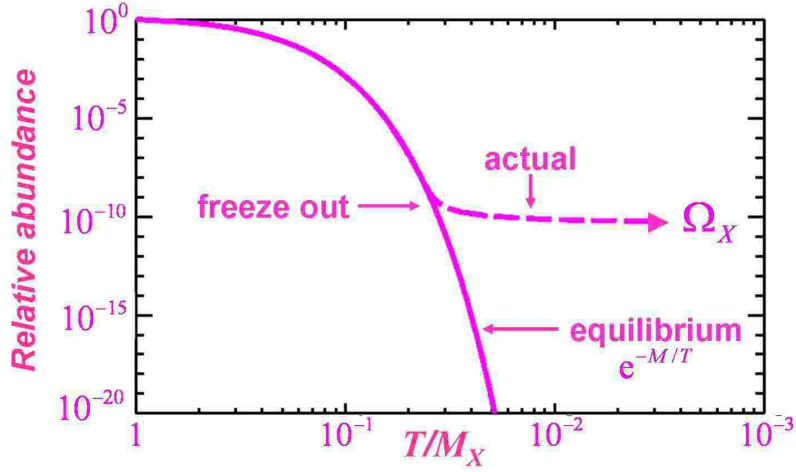}
\caption{A cartoon illustrating freezeout of a cold thermal relic.}
\label{freezeout}
\end{center}
\end{figure}

The first candidate I want to consider is a cold thermal relic.  Let's
call the particle $X$.  The idea behind a cold thermal relic is
illustrated in Fig.\ \ref{freezeout}.  The first assumption of a cold
thermal relic is that the particle was in local thermal equilibrium
(LTE) when the temperature was greater than the mass of the particle.
The second assumption is that there is no asymmetry between $X$ and
$\bar{X}$.\footnote{In some dark-matter models, the $X$ is a Majorana
particle, in which case $X\equiv\bar{X}$.}  The final assumption is
that the $X$ remains in LTE until temperatures drop below
$M_X$, and the $X$ becomes ``cold.''

With the above assumptions, for $T>M_X$ the $X$ should be about as
abundant as photons.  For $T<M_X$, the abundance of the $X$ relative
to photons is Boltzmann suppressed as long as the $X$ remains in LTE.

%%%%%%%%%%%%%%%%%%%%%%%%%%
%%%%%%%%%%%%%%%%%%%%
%%%%%%%%%%%%%%%%%%%%%
%%%%%%%%%%%%%%%%%%%%%%%%%%

But eventually the $X$ will ``freeze out'' of LTE.  As $T$ drops below
$M_X$, the particle becomes exponentially rare, so annihilation shuts
off.  Also, as the temperature drops below $M_X$ it becomes
exponentially unlikely for a collision in the plasma to have enough
center-of-momentum energy to create a $X\bar{X}$ pair.

The important feature of a cold thermal relic is that the more strongly a
particle interacts, the greater will be its annihilation (and creation) cross
section, the longer it will remain in equilibrium, and the \textit{lower} its
eventual freeze-out abundance will be.  So the weaker the interactions, the
larger will be the final abundance.

%%%%%%%%%%%%%%%%%%%%
%%%%%%%%%%%%%%%%%%%%
\begin{figure}
\begin{center}
\includegraphics[width=11cm]{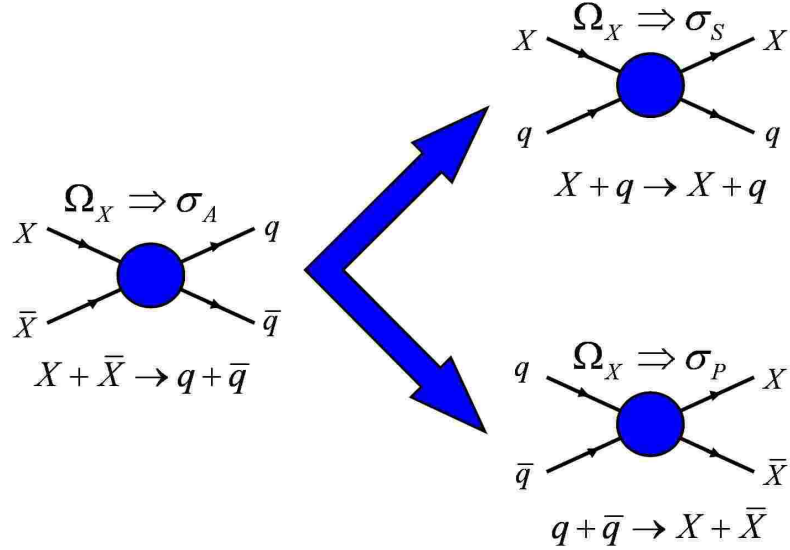}
\caption{A cartoon illustrating how knowledge of the annihilation cross section
tells us something about the scattering and production cross sections.}
\label{crossing}
\end{center}
\end{figure}
%%%%%%%%%%%%%%%%%%%%%
%%%%%%%%%%%%%%%%%%%%%

Detailed calculations of the freeze-out abundance yield the result
that the final abundance is proportional to the inverse of the
annihilation cross section, $\sigma_A$, and proportional to the
inverse of $M_X$.  Since the present contribution to $\Omega$ for the
$X$ is equal to $M_X$ times the present number density of $X$,
$\Omega_X \propto \sigma_A^{-1}$, and \textit{independent} of
$M_X$.\footnote{More exactly, to first approximation $\Omega_X$
depends on $M_X$ only through the dependence of $\sigma_A$ on $M_X$.}

Now if we want $\Omega_X$ to be the dark matter, we know the
approximate value of $\Omega_X$ required, so we know the approximate value of
$\sigma_A$.  Since it turns out to be of the order of the weak scale,
the $X$ is described as a weakly interacting massive particle, or
WIMP.  From now on, I will refer to a cold, thermal relic as a WIMP.

If we know the annihilation cross section, then we know something
about the scattering cross section and the production cross section.
This is illustrated in Fig.\ \ref{crossing} in the case that the WIMP
annihilates into $q\bar{q}$.

So if the WIMP has a weak-scale annihilation cross section, the
scattering cross sections should also be weak scale, and the
production cross section for \textrm{direct} production should also be
weak scale.

A weak-scale cross section is very interesting.  A cross section that
large should be within reach of \textit{direct detection} experiments,
where the relic WIMPs are detected by the small energy deposited in a
sensitive detector.  There are presently about a dozen experiments
that are underway or planned.

A weak-scale cross section also offers exciting possibilities for
\textit{indirect detection}.  The idea is that the WIMPs gather in the
center of galaxies, the Sun, or Earth, they annihilate, and produce a
signal that can be distinguished from anything produced by 
ordinary astrophysical
processes.  For instance, if the annihilation products include
high-energy neutrinos, positrons, or antiprotons, that could be a
signal for WIMPs.

Finally, if the annihilation cross section is weak scale, then the production
cross section should also be weak scale.  That should be within the range of
production and detection at accelerators with enough center-of-mass energy to
produce the WIMPs.  If this happens, then for the first time in the last 13.78
billion years, WIMPs would be produced in abundance.  This time, they would be
produced not as a result  of the tremendous temperatures of the big bang, but
rather as the result of human curiosity and ingenuity.

The favorite candidate for a WIMP is the neutralino of low-energy
SUSY.  The neutralino is a linear combination of gauginos (winos and
binos) and Higgsinos.  The mass and interactions of the lightest neutralino
depends on the particular linear combination of the gauginos and
Higgsinos, and its mass.

There are over one hundred new parameters in low-energy SUSY.  In
order to get some sort of handle on the problem, SUSY is usually
studied within some sort of ``constrained'' models.  Within these
constrained models, generally what is discovered is that SUSY models
consistent with accelerator data typically have too small an
annihilation cross section, which results in too large a value of
$\Omega_X$ (recall $\Omega_X \propto \sigma_A^{-1}$).  Therefore some
sort of cleverness/chicanery is required to increase the annihilation
cross section and give an acceptable $\Omega_X$.  The possibilities
include $s$-channel resonance through light $H$ and $Z$ poles,
co-annihilation with stops or staus, large $\tan\beta$
models\footnote{Here $\tan\beta$ is the ratio of vacuum expectation
values of the two Higgs of the model.} where annihilation occurs
through a broad $A$ resonance, or high values of the universal scalar
mass that makes the neutralino Higgsino-like so it can annihilate
easily into $W$ and $Z$ pairs.  Or, perhaps the true low-energy SUSY
model (if there is one) is unconstrained.

Today, direct detection experiments, indirect detection experiments,
and colliders are racing for discovery of the WIMP.  Imagine that by
2010 we have credible signals from all three.  A question to ask is,
\textit{how will we know we are seeing the same phenomenon?}  Let's
hope for this problem!

There area lot of opinions (i.e., papers) on this subject, let me
mention just three that are representative of the spread of opinion.
Arnowitt and Dutta \cite{dick} conclude that we will learn enough
about SUSY from the LHC to be confident that the particle discovered
in colliders is the same particle seen in direct and indirect
detection experiments. Baltz et al. \cite{ted} conclude that we will
not learn enough from the LHC, and we will require a ILC.  Finally,
Chung et al. \cite{dan} conclude that the answer depends on where in
parameter space the low-energy SUSY model lives.

As mentioned, the neutralino is everyone's favorite dark matter
candidate because of the rich physics possibilities it will provide.
It is fair to say that if dark matter is a complex natural phenomenon,
the neutralino is a simple, elegant, compelling explanation.  However,
that doesn't mean it is right.  As Tommy Gold once told me, ``For
every complex natural phenomenon, there is a simple, elegant,
compelling, \underline{\textit{wrong}} explanation.'' So I plead with
you to keep an open mind, and since the title of this course is
\textit{Searching for the `totally unexpected' in the LHC era,} and to
many people it would be totally unexpected that the neutralino is
\textit{not} the dark matter, let's keep an open mind and consider
some other possibilities.

First, let's consider the possibility that the dark matter is a cold thermal
relic other than a neutralino.  One possibility is that it is a Kaluza--Klein
(KK) excitation.  In 1984, Dick Slansky \cite{Slansky} and I pointed out that
since extra-dimensional theories have excited states corresponding to momentum
in extra dimensions, if there is some sort of conservation law keeping the
extra dimensions stable, the KK excitations would be thermally produced in the
early universe and appear as dark matter.  We only considered the possibility
that the size of extra dimensions are Planckian (size about $G^{-1/2}$), with
the result that there would be way too much dark matter.  Recently Servant and
Tait \cite{Geraldine}, and Cheng, Feng, and Matchev \cite{Konstantin}
independently returned to this scenario.  They greatly improved our analysis in
two ways: 1) They had a much better explanation of the origin of the parity
keeping the first excited mode stable (involving orbifolding the extra
dimension---that technology was not around in 1984), and 2) they imagined the
extra dimensions could be as large as the weak scale (which would have been
considered crazy in 1984).

In any case, detailed analysis suggests that the lightest KK mode
(presumably the KK photon) is an excellent dark matter candidate.
Furthermore, the collider signal for the KK excitations could be
confused with SUSY.  This scenario leads to very interesting
possibilities.

%%%%%%%%%%%%%%%%%%%%%%%%%%
%%%%%%%%%%%%%%%%%%%%%%%%%%
\subsection{Solitons}
%%%%%%%%%%%%%%%%%%%%%%%%%%
%%%%%%%%%%%%%%%%%%%%%%%%%%

So far I have assumed that the dark matter is an elementary particle. However,
there are other possibilities.  In particular, the dark matter may be in the
form of a soliton configuration known as a $Q$-ball or a non-topological
soliton.

If there is a scalar field with a conserved global charge $Q$, then
there are scalar field configurations that consist of a lump of
coherent scalar condensate of charge $Q$ in which the energy scales as
$Q^{3/4}$, so the soliton can not decay to $Q$ single-particle states.

$Q$-balls exist in the minimal supersymmetric standard model
\cite{alex1} where the scalar fields are squarks and sleptons, and the
conserved charge is baryon (or lepton) number.  The mass of the $Q$-ball
is $M\sim(1\textrm{ TeV})B^{3/4}$.  There is a penalty to pay in the
energy, so one has to go to very large $B$ to find a stable soliton;
in this case the minimum $B$ is $10^{12}$.

There are several mechanisms for producing solitons \cite{joshetal};
in this case, the most promising possibility is a fragmentation of an
Affleck-Dine condensate.  If this is the correct scenario it is
possible that the baryon asymmetry is also generated through the
fragmentation, so it is possible to relate the baryon asymmetry (hence
$\Omega_B$) to the dark matter density.

Note that solitons have a non-thermal origin. 

%%%%%%%%%%%%%%%%%%%%%%%%%%
%%%%%%%%%%%%%%%%%%%%%%%%%%
\subsection{Supermassive relics}
%%%%%%%%%%%%%%%%%%%%%%%%%%
%%%%%%%%%%%%%%%%%%%%%%%%%%

Recall that the present mass density for thermal relics is
proportional to $\sigma_A^{-1}$.  Therefore, there is a
\textit{minimum} annihilation cross section to ensure that
$\Omega_X<1$.

But for a particle of mass $M_X$, there is a maximum cross section for
each partial wave based on unitarity; roughly $\sigma_A<M_X^{-2}$.

Since there is a minimum annihilation cross section and a maximum
annihilation cross section (which depends on the mass), there is a
maximum mass for the thermal WIMP if $\Omega_X<1$.  The mass turns out
to be about 240 TeV \cite{gk}.

So if the dark matter has a mass in excess of 240 TeV, it must have a
nonthermal origin.  Solitons discussed in the previous subsection is
an example of a nonthermal relic.  My favorite example of a massive
nonthermal relic is the WIMPZILLA.  The most promising WIMPZILLA
production mechanism is gravitational production during inflation.  So
I will postpone the discussion of WIMPZILLAS until after I have said a
few words about inflation.

%%%%%%%%%%%%%%%%%%%%%%%%%%
%%%%%%%%%%%%%%%%%%%%%%%%%%
\subsection{Other possibilities}
%%%%%%%%%%%%%%%%%%%%%%%%%%
%%%%%%%%%%%%%%%%%%%%%%%%%%

Because of the lack of time I will not be able to cover other
possibilities.  I could have equally well discussed neutrinos,
gravitinos, axions, axion clusters, axinos, and so forth.

%%%%%%%%%%%%%%%%%%%%%%%%%%
%%%%%%%%%%%%%%%%%%%%%%%%%%
\section{Inflation}
%%%%%%%%%%%%%%%%%%%%%%%%%%
%%%%%%%%%%%%%%%%%%%%%%%%%%

The subject of inflation is well developed, but not well understood.  I won't
have time to discuss the basics of inflation: motivation, how inflation solves
classical cosmological problems like the age problem, the flatness problem, the
monopole problem, whether inflation is eternal or not, transplanckian issues,
and so on.

Here I will only say a few words about the origin of perturbations.  I choose
this because the measurements of the perturbation properties can tell us
something about inflation models.

%%%%%%%%%%%%%%%%%%%%
%%%%%%%%%%%%%%%%%%%%
\begin{figure}
\begin{center}
\includegraphics[width=11cm]{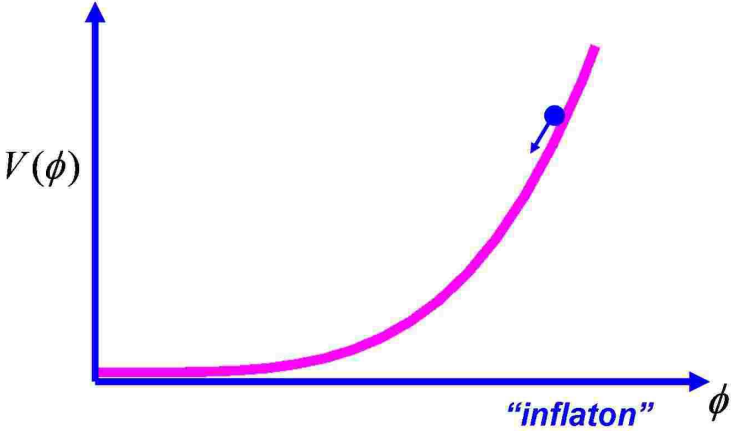}
\caption{A cartoon illustrating the classical dynamics of inflation.}
\label{inflation}
\end{center}
\end{figure}
%%%%%%%%%%%%%%%%%%%%%
%%%%%%%%%%%%%%%%%%%%%

The way we usually picture inflation is shown in Fig.\
\ref{inflation}. A scalar field, called the \textit{inflaton}, evolves
classically in an inflaton potential.  While the inflaton is displaced
from its minimum (here conveniently adjusted to have vanishing
potential energy at the minimum), there is an effective vacuum energy
$\rho_V=V(\phi)$ that dominates the energy density and drives a
quasi-de Sitter phase.

However, the classical picture is not the complete picture.  As first
pointed out by Schr\"{o}dinger in 1939, the expanding Universe leads
to particle creation.  I won't have time to go through the quantum
field theory arguments, but I will simply try to give a physical
motivation why the expanding universe leads to particle creation.

%%%%%%%%%%%%%%%%%%%%
%%%%%%%%%%%%%%%%%%%%
\begin{figure}
\begin{center}
\includegraphics[width=11cm]{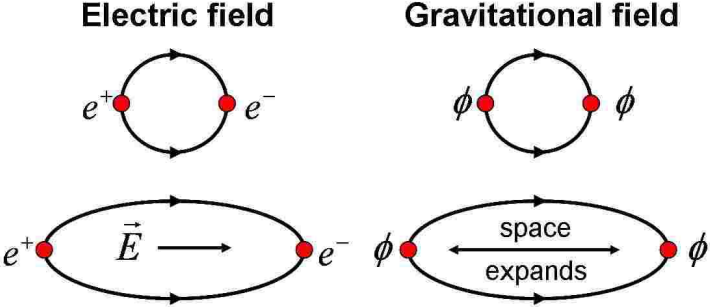}
\caption{A strong electromagnetic field affects vacuum fluctuations,
and if sufficiently strong, can lead to particle production.  The
expanding Universe can also rip apart vacuum fluctuations and lead to
particle creation.}
\label{production}
\end{center}
\end{figure}
%%%%%%%%%%%%%%%%%%%%%
%%%%%%%%%%%%%%%%%%%%%

As illustrated in Fig.\ \ref{production}, it is well known that
particle production is possible in a sufficiently strong
electromagnetic field.  If the energy gained in acceleration of an
$e^+e^-$ pair in a distance of an electron Compton wavelength exceeds
the electron rest mass, then pair creation can occur.  It is as if the
strong electromagnetic field rips $e^+e^-$ pairs out of the vacuum
fluctuations.

One can picture the expansion of the Universe having the same effect.
Particles and antiparticles come out of the vacuum, get caught in the
expansion of space, and are ripped out of the virtual sea and turned
to real particles.

The potential energy of the inflaton field is the energy of the
zero-momentum mode of the field.  By definition, the zero-momentum
mode is homogeneous and isotropic.  As particles of the inflaton field
are created due to expansion of the Universe, they are created with
non-zero momentum.  If there are non-zero momentum modes of the
inflaton field, the inflaton field can not be perfectly homogeneous
and isotropic.

Since the inflaton field dominates the energy density, the
inhomogeneities in the inflaton field lead to metric fluctuations.
The metric fluctuations can be divided into fluctuations known as scalar
fluctuations, which are what we usually refer to as density
fluctuations, and tensor fluctuations, which are equivalent to a
gravitational wave background.

The scalar and tensor perturbation spectra are a function of the
expansion rate during inflation, and how the expansion rate changes
during inflation.  Since the inflaton potential determines the
expansion rate, the inflaton potential will determine the perturbation
spectra.

%%%%%%%%%%%%%%%%%%%%
%%%%%%%%%%%%%%%%%%%%
\begin{figure}
\begin{center}
\includegraphics[width=11cm]{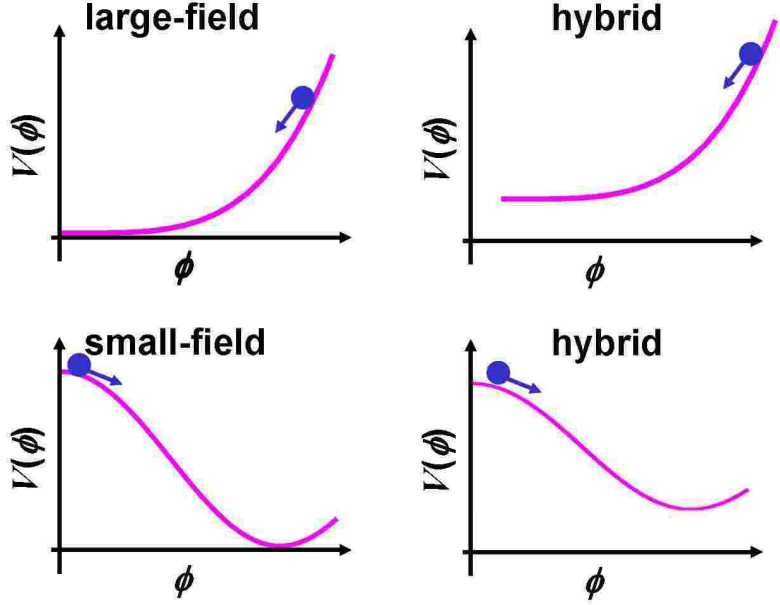}
\caption{Classes of inflationary models.}
\label{zoo}
\end{center}
\end{figure}
%%%%%%%%%%%%%%%%%%%%%
%%%%%%%%%%%%%%%%%%%%%

Some basic classes of inflaton models are illustrated in Fig.\
\ref{zoo}.  Large-field models have positive second derivative,
small-field models have negative second derivative, hybrid models can
have positive or negative second derivative, but when inflation ends
the vacuum energy is not zero, and there must be a second field required to
describe the non-inflationary evolution to the true ground state.

The scalar and tensor perturbation amplitudes are in general functions
of wavelength (or wavenumber $k$).  It is usually quite accurate to
parametrize the spectra in terms of amplitudes and spectral indices
as described in Table \ref{spectra}.

%%%%%%%%%%%%%%%%%%%%%%%%%%
%%%%%%%%%%%%%%%%%%%%%%%%%%
\begin{table}
\caption{\label{spectra} Parametrization of scalar and tensor perturbation
spectra.  Here, $k_*$ is some conveniently chosen wavenumber.  }
\begin{ruledtabular}
\begin{tabular}{lcr}
      & scalar perturbations  & tensor perturbations   \\ 
      & (density perturbations) & (gravitational waves) \\ \hline
amplitude at $k=k_*$ & $P_S(k_*)$ & $ P_T(k_*) $\\
spectral index       & $n_S \equiv \Frac{d\ln P_S(k_*)}{d\ln k}$ &
$n_T \equiv \Frac{d\ln P_T(k_*)}{d\ln k}$ \\
running of the spectral index  & $n_S^\prime \equiv \Frac{dn_S}{d\ln k}$ &
$n_T^\prime \equiv \Frac{dn_T}{d\ln k}$ \\
\end{tabular}
\end{ruledtabular}
\end{table}
%%%%%%%%%%%%%%%%%%%%%%%%%%
%%%%%%%%%%%%%%%%%%%%%%%%%%

There are a few things we have learned from attempts to build
inflation models.
\begin{enumerate}
\item The inflaton potential must be ``flat'' in the sense that its
derivative must be unusually small.
\item While it is always easy to write a potential that is flat, it
must remain flat when one includes radiative corrections.  This
presumably suggests some (possibly approximate) symmetry.
\item Although it has often been said that SUSY can come to the
rescue, in practice it is not so easy to implement \cite{toni}.
\item Many attractive SUSY models give $V(\phi) \sim A+\ln \phi$
hybrid models.
\item There are no general predictions, but in many models we find:
\begin{enumerate}
\item Ratio of tensor to scalar amplitudes: 
$r\sim P_T(k_*)/P_S(k_*)\sim (\textrm{small})$.
\item Scalar spectral index: $\left | n_S-1 \right| \sim (\textrm{small})$.
\item Tensor spectral index: $\left | n_T \right| \sim (\textrm{small})$.
\item Running of $n_S$:  $\left| n_S^\prime \right| \sim (\textrm{small})^2$.
\item Running of $n_T$:  $\left| n_T^\prime \right| \sim (\textrm{small})^2$.
\end{enumerate}
\item We need experimental guidance.
\end{enumerate}

\begin{figure}
\begin{center}
\includegraphics[width=11cm]{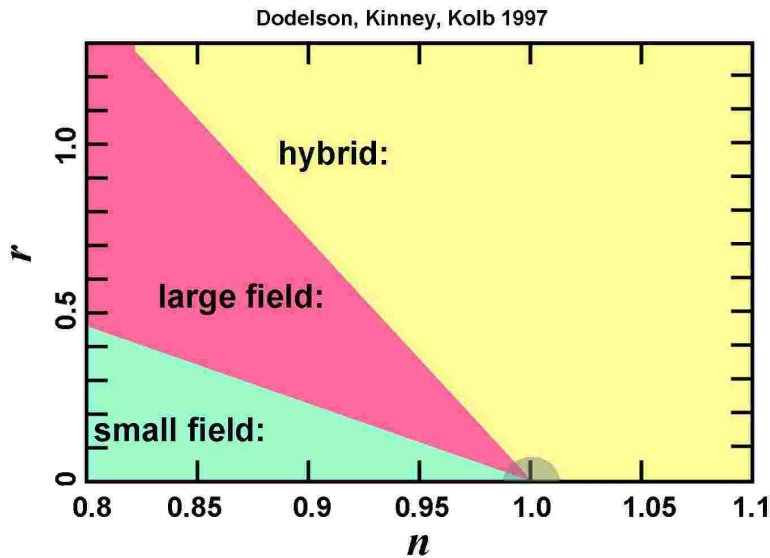}
\caption{The observational plane of $r \propto P_T(k_*)/P_S(k_*)$ and $n=n_S$. 
Notice the convergence of models at $n=1$ and $r=0$.}
\label{rn1}
\end{center}
\end{figure}

For phenomenological guidance, it is now traditional to present the
observational results on $n_S$ and the ratio of tensor to scalar
perturbations on the same graph \cite{will}.  The utility of this
approach can be seen in Fig.\ \ref{rn1}.  To a first approximation,
small-field, large-field, and hybrid models populate different regions
of the $n_S$--$r$ plane.
%%%%%%%%%%%%%%%%%%%%
%%%%%%%%%%%%%%%%%%%%
%%%%%%%%%%%%%%%%%%%%%
%%%%%%%%%%%%%%%%%%%%%

\begin{figure}
\begin{center}
\includegraphics[width=16cm]{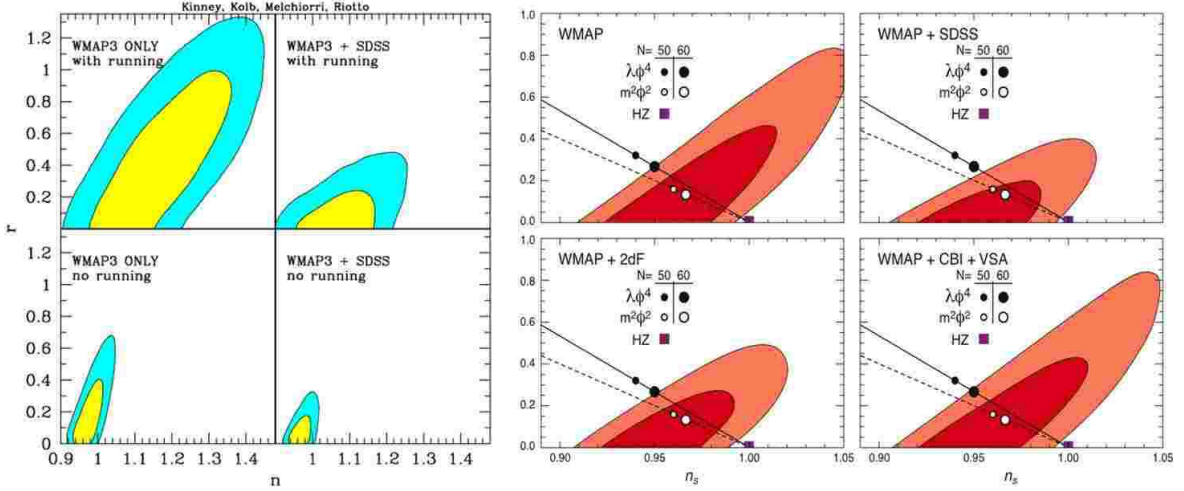}
\caption{Two analysis showing the experimental situation The analysis on the 
left is from Ref.\ \cite{ale} and on the right from \cite{wmap3}. }
\label{rn2}
\end{center}
\end{figure}

The experimental situation is shown in Fig.\ \ref{rn2}.  If one
analyzes only the WMAP three-year data set, and takes a prior that
there is no running of the spectral index, the Harrison--Zel'dovich
spectrum ($n_S=1$, $n_S^\prime=0$) is ruled out at approximately the
$90\%$ confidence level.  The result is little changed if one adds
additional data sets.  If one allows running of the scalar spectral
index, then a slightly blue ($n>1$) spectrum is preferred.  My
conclusion is that one can not make a strong statement about whether
the Harrison--Zel'dovich spectrum is ruled out or not.  It is also
premature to draw any conclusion regarding phenomenology of inflation
models.

%%%%%%%%%%%%%%%%%%%%
%%%%%%%%%%%%%%%%%%%%
%%%%%%%%%%%%%%%%%%%%%
%%%%%%%%%%%%%%%%%%%%%

%%%%%%%%%%%%%%%%%%%%%%%%%%
%%%%%%%%%%%%%%%%%%%%%%%%%%
\subsection{Wimpzillas} 
%%%%%%%%%%%%%%%%%%%%%%%%%%
%%%%%%%%%%%%%%%%%%%%%%%%%%

Now let us return to the possibility of producing non-thermal dark
matter during inflation.  Particles of all types (at least those that
are not conformally coupled) are produced during inflation.  If there
is a massive stable particle species, they will also be produced during
inflation and be around today.  Two groups \cite{ckr,kt} proposed this
could be the source of dark matter.

Detailed calculations show that that if there is a stable particle
with mass comparable to the inflaton mass, they would be produced in
the proper abundance to be the dark matter.  Since the mass of the
inflaton is expected to be $10^{11}$ or $10^{12}$ GeV, the dark matter
particle would be supermassive---much more massive than a WIMP can
be---hence, it is known as a WIMPZILLA.

Unlike thermal relics, where the contribution to $\Omega$ depends on
the annihilation cross section and independent of the mass, the
WIMPZILLA contribution to $\Omega$ is independent of the interactions
and depends only on the mass

%%%%%%%%%%%%%%%%%%%%%%%%%%
%%%%%%%%%%%%%%%%%%%%%%%%%%
\acknowledgments{It is a pleasure to thank Profs.\ A.\ Zichichi and
G.\ 't Hooft for the invitation to lecture at the school, and the
students of the school for many questions about the subjects presented
here.}
%%%%%%%%%%%%%%%%%%%%%%%%%%
%%%%%%%%%%%%%%%%%%%%%%%%%%

%%%%%%%%%%%%%%%%%%%%%%%%%%
%%%%%%%%%%%%%%%%%%%%%%%%%%

%%%%%%%%%%%%%%%%%%%%%%%%%%
%%%%%%%%%%%%%%%%%%%%%%%%%%

%%%%%%%%%%%%%%%%%%%%%%%%%%
%%%%%%%%%%%%%%%%%%%%%%%%%%
%%%%%%%%%%%%%%%%%%%%%%%%%%
\end{document}